\documentclass[12pt,aps,amsmath,amssymb]{revtex4} % SINGLE SPACED !!!!!!

\usepackage{graphicx}% Include figure files
\usepackage{dcolumn}% Align table columns on decimal point
\usepackage{bm}% bold math
\usepackage{setspace}
\usepackage{color}

\begin{document}
\begin{singlespace}

\title{Accuracy of direct gradient sensing by cell-surface receptors}

\author{Robert G.~Endres$^{1,2}$, Ned S.~Wingreen$^{3}$}
 \affiliation{
$^{1}${Division of Molecular Biosciences, Imperial College London, London SW7 2AZ, United Kingdom} \\
$^{2}${Centre for Integrated Systems Biology at Imperial College, Imperial College London, 
London SW7 2AZ, United Kingdom},\\
$^{3}$Department of Molecular Biology, Princeton University, Princeton, NJ 08544-1014.\\
}

\date{\today}

\begin{abstract}
Chemotactic cells of eukaryotic organisms are able to accurately sense shallow
chemical concentration gradients using cell-surface receptors. This sensing 
ability is remarkable as cells must be able to spatially resolve small
fractional differences in the numbers of
particles randomly arriving at cell-surface receptors by diffusion. 
An additional challenge and source of uncertainty is that particles,
once bound and released, may rebind the same or a different 
receptor, which adds to noise without providing any new information about 
the environment. We recently derived the 
fundamental physical limits of gradient sensing 
using a simple spherical-cell model, but not including explicit particle-receptor 
kinetics. 
Here, we use a method based on the fluctuation-dissipation theorem (FDT) 
to calculate the accuracy of gradient sensing by realistic receptors. 
We derive analytical results for two receptors, as well as two coaxial 
rings of receptors, {\it e.g.} one at each cell pole.
For realistic receptors, we find that particle rebinding lowers the 
accuracy of gradient sensing, in line with our previous results.
\end{abstract}

\maketitle

\section{Introduction}
Cells are able to sense gradients of chemical concentration 
with extremely high sensitivity and accuracy. This is done either directly, by measuring spatial gradients
across the cell diameter, or indirectly, by temporally sensing gradients while 
moving \cite{dusenbery98}. In temporal sensing, a cell modifies its swimming behavior according to whether 
a chemical concentration is rising or falling in time \cite{berg99}. 
This mode of sensing is typical of small, fast moving bacteria such as 
{\it Escherichia coli}, which can respond to changes in concentration as low as 
3.2 nM of the attractant aspartate \cite{manson03}. In contrast, direct spatial sensing is 
prevalent among larger, single-celled eukaryotic organisms such as the slime mold 
{\it Dictyostelium discoideum} (Dicty) and the budding yeast {\it Saccharomyces cerevisiae} 
\cite{arkowitz99,manahan04}. Dicty cells are able to sense a concentration difference 
of only 1-5\% across the cell \cite{mato75}, corresponding to a difference in receptor 
occupancy between front and back of only 5 receptors \cite{haastert07}. Spatial 
sensing is also performed by cells of the immune system including 
neutrophils and lymphocytes \cite{zigmond77}, as well as by growing synaptic 
cells and tumor cells. Interestingly, a direct spatial mode of sensing has also been demonstrated
for the large oxygen-sensing bacterium {\it Thiovulum majus} \cite{thar03}, 
indicating that direct gradient sensing is widespread among the different 
kingdoms of life.

There has been great progress in understanding the limits of
concentration sensing in bacteria such as {\it E. coli}, 
pioneered by Berg \cite{berg77,berg99,sourjik2002}, and in understanding
the origins of sensitivity in the underlying signaling network, pioneered by
Bray \cite{bray98,duke99,bray02,bray07}, and followed
by others \cite{SB2004,mello05,keymer06,endres06,bialek08,hansen08,endres_MSB08}. 
By contrast, very little is known about what determines the accuracy of direct gradient
sensing by eukaryotic cells. Recently, we derived
the fundamental physical limits of direct gradient sensing, where the accuracy is 
limited by the random arrival of particles at the cell surface due to 
diffusion \cite{endres_PNAS08}.
We used as models a perfectly absorbing sphere and a perfectly
monitoring sphere (\`a la Berg and Purcell \cite{berg77}).
In these two models, gradients are inferred from the positions of particles absorbed 
on the surface of a sphere or the
positions of freely diffusing particles inside a spherical
volume, respectively. The latter case simulates rebinding of particles, as particles can
enter and exit the spherical volume freely. In comparison, for the 
perfectly absorbing sphere,  previously observed particles are never remeasured.
As a result, we found that the perfectly absorbing sphere is superior to the
perfectly monitoring sphere, both for concentration and gradient sensing (Table I).

The superiority of the absorbing sphere may help explain the presence at the surfaces of 
cells of signal degrading enzymes, such as PDE for cAMP in Dicty and BAR1 for mating factor 
$\alpha$ in {\it S. cerevisiae}. Those surface enzymes could reduce or eliminate rebinding 
(and therefore remeasurement) of the same signal molecule.
Quantitatively, our theory compares favorably to recent 
measurements of Dicty moving up shallow cAMP gradients  \cite{haastert07}, 
suggesting that these cells operate near the physical limits of gradient detection.

While our recent models of the absorbing and monitoring spheres
allowed us to derive the fundamental limit of gradient sensing, the 
models neglect the details of biochemical reactions, such as particle-receptor binding
and downstream signaling events,
which might further increase measurement uncertainty. To study the
effects of particle-receptor binding, we here extend a formalism for
the uncertainty of concentration 
sensing recently developed by Bialek and Setayeshgar \cite{bialek05,bialek08},
to the case of gradient sensing. This formalism uses the 
fluctuation-dissipation theorem to infer the fluctuations of the receptor occupancy (and
hence the accuracy of concentration sensing) from the linear response of the average
receptor occupancy to changes in receptor binding free energies. The effect of 
particle rebinding is included by coupling particle-receptor binding to the diffusion equation \cite{bialek05}, 
leading to correlations in time among the receptors.  
We report analytical results for two receptors (Fig. 1), as well as two coaxial 
rings of receptors, {\it e.g.} one at each cell pole (Fig. 2). By assuming
diffusion-limited particle binding to the receptors, we are able 
to directly compare to our previous model for the fundamental limits
of gradient sensing.
For realistic receptors, we find that particle rebinding lowers the 
accuracy of gradient sensing in line with our previous results for
the absorbing and monitoring spheres (Table I).

\begin{table}[t]
%\begin{minipage}[t]{7.5cm}
\linespread{1.5}
%\large
\begin{tabular}{c|c|c|c}
               Measurement uncertainty   & Perfect absorber  &  Perfect monitor  & Ratio absorber/monitor \\
\hline

Concentration: $\frac{\langle(\delta c)^2\rangle}{c_0^2}$           & $\frac{1}{4\pi Dac_0T}$  & $\frac{3}{5\pi Dac_0T}$ \cite{berg77} & $\frac{12}{5}${\normalsize$=2.4$}   \\
Gradient: $\frac{\langle(\delta c_{\vec r})^2\rangle}{(c_0/a)^2}$ & $\frac{1}{4\pi Dac_0 T}$ & $\frac{15}{7\pi Dac_0 T}$ & $\frac{60}{7}${\normalsize$\approx 8.6$} \\
\end{tabular}
\linespread{1.0}
\caption{\label{tab:formulas} 
Uncertainties in measured concentration and concentration gradient for two idealized cell 
models: a perfectly absorbing sphere (second column) and a perfectly monitoring sphere 
(third column). Also provided is the ratio of the uncertainties of the absorber and monitor. 
Parameters: diffusion constant $D$, radius of sphere $a$, averaging time $T$, and average
chemical concentration $c_0$. Table reproduced from Ref. \cite{endres_PNAS08}.}
%\end{minipage}
\end{table}

\begin{figure}             
\includegraphics[width=7.5cm,angle=0]{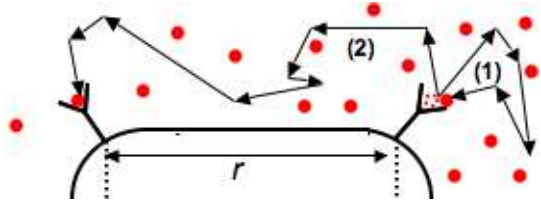}
\caption{Model for gradient sensing by two individual cell-surface receptors. 
Particles diffuse through the medium and randomly bind to and unbind
from receptors, {\it e.g.} see the two sample paths indicated by black arrows.
Particles may rebind the same (1) or a different (2) receptor. 
The receptor separation is given by $r$.}
\label{fig:fig1}
\end{figure}

\section{Methods}
Bialek and Setayeshgar \cite{bialek05} previously presented a method based on the
fluctuation-dissipation theorem (FDT) \cite{kubo66} to calculate the accuracy of measurement of
chemical concentration by receptors. We extend their method to calculate the accuracy of measurement of 
concentration gradients, and derive analytical results for (1) two receptors and (2) two
coaxial rings of receptors, {\it e.g.} one at each cell pole. We start our derivation by
considering an arbitrary number $m$ of receptors.

The kinetics of the ensemble-average occupancy $n_j(t)$ of receptor $j$ due to binding and unbinding of chemical ligands 
at the local concentration $c(\vec x_j,t)$ is given by 
\begin{equation}
\frac{dn_j(t)}{dt}=k_+c(\vec{x}_j,t)[1-n_j(t)]-k_-n_j(t).
\end{equation}
Linearization about the mean steady-state occupancy $\bar n_j=k_+\bar c_j/(k_+\bar c_j+k_-)$ at concentration 
$\bar c_j=\bar c(\vec x_j,t)$ gives
\begin{equation}
\frac{d(\delta n_j(t))}{dt}=-(k_+\bar c_j+k_-)\delta n_j+\bar c_j(1-\bar n_j)\delta k_+
-\bar n_j\delta k_-+k_+(1-\bar n_j)\delta c_j.\label{eq:dn}
\end{equation}
By thermodynamics, the ratio of binding and unbinding rates is related to the free-energy difference $F_j$ between the unbound
and bound states of the receptor according to
\begin{equation}
\frac{k_+\bar c_j}{k_-}=\exp\left(\frac{F_j}{k_BT}\right).
\end{equation}
Variations of the rate constants in Eq. \ref{eq:dn} are equivalent to a variation or external perturbation 
of this receptor free-energy difference
\begin{equation}
\frac{\delta F_j}{k_BT}=\frac{\delta k_+}{k_+}-\frac{\delta k_-}{k_-}.\label{eq:dk}
\end{equation}
Combining Eqs. \ref{eq:dn} and \ref{eq:dk}, one obtains after Fourier transforming in time to obtain a
frequency representation
\begin{equation}
-i\omega\delta\hat n_j(\omega)=-(k_+\bar c_j+k_-)\delta \hat n_j(\omega)+\frac{k_+(1-\bar n_j)\bar c_j}{k_BT}
\delta\hat F_j(\omega)+k_+(1-\bar n_j)\delta \hat c_j(\omega),\label{eq:FTn}
\end{equation}
where, {\it e.g.}, $\delta \hat n_j(\omega)$ describes the variations in the receptor occupancy as a function of
frequency $\omega$. The dependence on $\delta \hat c_j(\omega)$ can be eliminated by using 
the (linearized) diffusion equation \cite{bialek05} in Eq. \ref{eq:FTn}
\begin{equation}
\frac{\partial (\delta c(\vec{x},t))}{\partial t}=D\nabla^2\delta c(\vec{x},t)-
\sum_{l=1}^m\delta(\vec{x}-\vec{x_l})\frac{d(\delta n_l(t))}{dt}.\label{eq:diff}
\end{equation}
Using the Fourier transforms
\begin{eqnarray}
\delta c(\vec x,t)&=&\int\frac{d\omega}{2\pi}\int\frac{d^3k}{(2\pi)^3}e^{i(\vec k\vec x-\omega t)}\delta\hat c(\omega,\vec k)\\
\delta (\vec x-\vec x_l)&=&\int \frac{d^3k}{(2\pi)^3}e^{i\vec k(\vec x-\vec x_l)-k/\Lambda}\label{eq:df}\\
\delta n_l(t)&=&\int\frac{d\omega}{2\pi}e^{-i\omega t}\delta \hat n_l(\omega),
\end{eqnarray}
where $k=|\vec k|$ and where we have introduced a convergence factor $\Lambda\gtrsim 0$ in Eq. \ref{eq:df} 
to regulate the $\delta$ function (effectively assigning a size scale $\sim 1/\Lambda$ to the receptor),
Eq. \ref{eq:diff} yields
\begin{equation}
\delta\hat c(\omega,\vec k)=\frac{i\omega}{Dk^2-i\omega}\sum_{l=1}^me^{-i\vec k\vec x_l-k/\Lambda}\delta \hat n_l(\omega).
\end{equation}
Inverting the spatial Fourier transform back into real space, one obtains the concentration fluctuations
at the locations of the receptors in terms of the occupancy fluctuations
\begin{eqnarray}
\delta \hat c(\vec x_j,\omega)&=&
i\omega \sum_{l=1}^m\delta \hat n_l(\omega)\int \frac{d^3k}{(2\pi)^3}\frac{e^{i\vec k(\vec x_j-\vec x_l)-k/\Lambda}}{Dk^2-i\omega}\\
&=&i\omega\left[\delta\hat n_j(\omega)\int\frac{d^3k}{(2\pi)^3}\frac{e^{-k/\Lambda}}{Dk^2-i\omega}+\sum_{l\neq j}^m\delta \hat n_l(\omega)
\int\frac{d^3k}{(2\pi)^3}\frac{e^{i\vec k(\vec x_j-\vec x_l)}}{Dk^2-i\omega}\right]\\
&=&\frac{i\omega}{2\pi^2}\left[\delta \hat n_j(\omega)\int_0^{\Lambda}\frac{k^2\,dk}{Dk^2-i\omega}+
\sum_{l\neq j}^m\frac{\delta \hat n_l(\omega)}{|\vec{x}_j-\vec{x}_l|}\int_0^{\infty}\frac{k\sin(k|\vec{x}_j-\vec{x}_l|)}{Dk^2-i\omega}dk\right].\label{eq:dcx}
\end{eqnarray}
The cut-off $\Lambda=\pi/s$, which accounts for the physical dimensions of
the receptor, is used in Eq. \ref{eq:dcx} to set an upper limit of integration \cite{bialek05}.

Following Bialek and Setayeshgar \cite{bialek05}, 
we imagine that the mechanism that reads out the receptor occupancy averages over a time $\tau$ 
long compared to the correlation time between binding and unbinding events of a receptor (see below). 
In this case, we can apply the low-frequency limit ($\omega<\!\!<D\Lambda^2$). Using the expression
\begin{equation}
\lim_{\omega\rightarrow 0}\frac{1}{v\mp i\omega}=\frac{P}{v}\pm i\pi\delta(0),
\end{equation}
where $v$ is real and $P$ is the principal value of the associated integral, Eq. \ref{eq:dcx} becomes
\begin{equation}
\delta \hat c(\vec{x_j},\omega)=\frac{i\omega\Lambda}{2\pi^2D}\delta \hat n_j(\omega)+
\frac{i\omega}{4\pi D}\sum_{l\neq j}^m\frac{\delta \hat n_l(\omega)}{|\vec{x}_j-\vec{x}_l|}.\label{eq:dcx2}
\end{equation}
Inserting Eq. \ref{eq:dcx2} for $\delta \hat c_j(\vec x,\omega)$ in Eq. \ref{eq:FTn} yields
\begin{eqnarray}
-i\omega\delta\hat n_j&=&\left[k_+(1-\bar n_j)\frac{i\omega\Lambda}{2\pi^2D}-(k_+\bar c_j+k_-)\right]\delta\hat n_j\nonumber\\
&&+k_+(1-\bar n_j)\frac{i\omega}{4\pi D}\sum_{l\neq j}^m\frac{\delta\hat n_l}{|\vec x_j-\vec x_l|}
+\frac{k_+(1-\bar n_j)\bar c_j}{k_BT}\delta\hat F_j,\label{eq:dni}
\end{eqnarray}
which depends only on the $\delta \hat n_j(\omega)$ and their conjugate variables $\delta \hat F_j(\omega)$.
These $m$ equations describe how deterministic frequency-dependent changes in the 
free-energy differences $\delta \hat F_j(\omega)$ affect the frequency-dependent occupancies 
$\delta \hat n_j(\omega)$ of all the receptors. \\

\begin{figure}             
\includegraphics[width=7.5cm,angle=0]{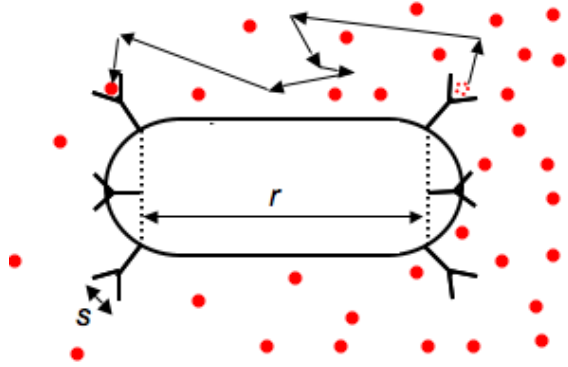}
\caption{Model for gradient sensing by two polar rings of surface receptors. 
Particles diffuse through the medium and randomly bind to and unbind from receptors.
Particles may rebind the same or different receptors ({\it e.g.}, path of black arrows). 
Rings of receptors are separated by distance $r$. Also shown is receptor dimension $s$.}
\label{fig:fig2}
\end{figure}

\noindent{\bf Two receptors}\\ 
We first consider two identical receptors (Fig. 1) for which Eq. \ref{eq:dni} simplifies to two coupled linear equations 
\begin{eqnarray}
-i\omega\delta\hat n_1&=&\left[k_+(1-\bar n_1)\frac{i\omega\Lambda}{2\pi^2D}-(k_+\bar c_1+k_-)\right]\delta\hat n_1\nonumber\\
&&+k_+(1-\bar n_1)\frac{i\omega}{4\pi D}\frac{\delta\hat n_2}{|\vec x_1-\vec x_2|}+\frac{k_+(1-\bar n_1)\bar c_1}{k_BT}\delta\hat F_1\\
-i\omega\delta\hat n_2&=&\left[k_+(1-\bar n_2)\frac{i\omega\Lambda}{2\pi^2D}-(k_+\bar c_2+k_-)\right]\delta\hat n_2\nonumber\\
&&+k_+(1-\bar n_2)\frac{i\omega}{4\pi D}\frac{\delta\hat n_1}{|\vec x_1-\vec x_2|}+\frac{k_+(1-\bar n_2)\bar c_2}{k_BT}\delta\hat F_2,
\end{eqnarray}
In matrix form these equations can be written in terms of a complex susceptibility $\chi$ as
\begin{equation}
\begin{pmatrix}
\delta\hat F_1 \\
\delta\hat F_2 
\end{pmatrix}
=\chi\cdot
\begin{pmatrix}
\delta\hat n_1 \\
\delta\hat n_2 
\end{pmatrix}
=k_BT
\begin{pmatrix}
\frac{k_+\bar c_1+k_--i\omega(1+\Sigma_1)}{k_+(1-\bar n_1)\bar c_1} & \frac{-i\omega}{4\pi Dr \bar c_1}\\
\frac{-i\omega}{4\pi D r \bar c_2} & \frac{k_+\bar c_2+k_--i\omega(1+\Sigma_2)}{k_+(1-\bar n_2)\bar c_2}
\end{pmatrix}
\cdot
\begin{pmatrix}
\delta\hat n_1 \\
\delta\hat n_2 
\end{pmatrix},\label{eq:eX}
\end{equation}
where $r=|\vec x_1-\vec x_2|$ and $\Sigma_i=\frac{k_+(1-\bar n_i)}{2\pi Ds}$.
The inverse susceptibility $\chi^{-1}$ relates changes in $\delta \hat F_i$ to changes in the occupancies $\delta \hat n_i$, {\it i.e.}

\begin{equation}
\begin{pmatrix}
\delta\hat n_1 \\
\delta\hat n_2 
\end{pmatrix}
=\chi^{-1}\cdot
\begin{pmatrix}
\delta\hat F_1 \\
\delta\hat F_2 
\end{pmatrix}
=
\begin{pmatrix}
\frac{\partial\hat n_1}{\partial\hat F_1} & \frac{\partial\hat n_1}{\partial\hat F_2} \\
\frac{\partial\hat n_2}{\partial\hat F_1} & \frac{\partial\hat n_2}{\partial\hat F_2}
\end{pmatrix}
\cdot
\begin{pmatrix}
\delta\hat F_1 \\
\delta\hat F_2 
\end{pmatrix}.
\end{equation}

According to the fluctuation-dissipation theorem \cite{bialek05}, the noise power spectra of the 
occupancies $S_{n_jn_l}(\omega)=\langle\delta \hat n_j(\omega)\delta \hat n_l^*(\omega)\rangle$ can be 
calculated in the low-frequency limit from the above deterministic response functions, 
obtained by inverting the matrix $\chi$ in Eq. \ref{eq:eX},
\begin{eqnarray}
S_{n_1n_1}(\omega\rightarrow 0)=\lim_{\omega \rightarrow 0}\frac{2k_BT}{\omega}\text{Im}\left(\frac{\partial \hat n_1}{\partial \hat F_1}\right)
&=&\frac{2k_+(1-\bar n_1)\bar c_1(1+\Sigma_1)}{(k_+\bar c_1+k_-)^2}\nonumber\\
&=&2\langle(\delta n_1)^2\rangle_\text{tot}\tau_{c_1}(1+\Sigma_1)\label{S11}\\
S_{n_2n_2}(\omega\rightarrow 0)=\lim_{\omega \rightarrow 0}\frac{2k_BT}{\omega}\text{Im}\left(\frac{\partial \hat n_2}{\partial \hat F_2}\right)
&=&\frac{2k_+(1-\bar n_2)\bar c_2(1+\Sigma_2)}{(k_+\bar c_2+k_-)^2}\nonumber\\
&=&2\langle(\delta n_2)^2\rangle_\text{tot}\tau_{c_2}(1+\Sigma_2)\label{S22}\\
S_{n_1n_2}(\omega\rightarrow 0)=\lim_{\omega \rightarrow 0}\frac{2k_BT}{\omega}\text{Im}\left(\frac{\partial \hat n_1}{\partial \hat F_2}\right)
&=&\frac{k_+^2(1-\bar n_1)(1-\bar n_2)\bar c_2}{2\pi D r(k_+\bar c_1+k_-)(k_+\bar c_2+k_-)}\nonumber\\
&=&\frac{\langle(\delta n_1)^2\rangle_\text{tot}\langle(\delta n_2)^2\rangle_\text{tot}}{2\pi Dr\bar c_1}\label{S12}\\
S_{n_2n_1}(\omega\rightarrow 0)=\lim_{\omega \rightarrow 0}\frac{2k_BT}{\omega}\text{Im}\left(\frac{\partial \hat n_2}{\partial \hat F_1}\right)
&=&\frac{k_+^2(1-\bar n_1)(1-\bar n_2)\bar c_1}{2\pi D r(k_+\bar c_1+k_-)(k_+\bar c_2+k_-)}\nonumber\\
&=&\frac{\langle(\delta n_1)^2\rangle_\text{tot}\langle(\delta n_2)^2\rangle_\text{tot}}{2\pi Dr\bar c_2}\label{S21},
\end{eqnarray}
where $\langle(\delta n_j)^2\rangle_\text{tot}=\bar n_j(1-\bar n_j)$ is the total variance in each
occupancy and $\tau_{c_j}=(k_+\bar c_j+k_-)^{-1}$ is the correlation time for receptor occupancy in the 
absence of rebinding. The correlation time $\tau_{c_j}$ is modified when rebinding is included via
coupling to particle diffusion so that the effective correlation time for receptor occupancy increases 
to $\tau_{c_j}(1+\Sigma_j)$ as shown by the above equations. For an equilibrium system with 
time-reversal symmetry, the power spectra are necessarily symmetric in the zero-frequency limit, 
{\it i.e.} $S_{n_1n_2}(0)=S_{n_2n_1}(0)$. However, $S_{n_1n_2}(0)$ (Eq. \ref{S12}) and $S_{n_2n_1}(0)$ (Eq. \ref{S21}) 
are different depending on $\bar c_1$ and $\bar c_2$, respectively. Hence, our approach based on the 
fluctuation-dissipation theorem is only valid for shallow gradients with $\bar c_1\approx\bar c_2$.  

To determine the uncertainty of gradient sensing, we need to know the  uncertainty in measuring a concentration difference.
This uncertainty can be obtained from the noise power spectra for ligand binding and unbinding (Eqs. \ref{S11}-\ref{S21}). 
Starting with the expression $\bar n=\bar c/(\bar c+K_D)$ for the average occupancy in terms of 
the average concentration, where $K_D=k_-/k_+$, we obtain the following relation
\begin{equation}
\delta c_j=\frac{(\bar c_j+K_D)^2}{K_D}\delta n_j=\frac{K_D}{(1-\bar n_j)^2}\delta n_j=\frac{k_-}{k_+(1-\bar n_j)^2}\delta n_j,\label{eq:dc_n}
\end{equation}
which expresses how a fluctuation in the occupancy of receptor $j$ affects the cell's best estimate of the local concentration.
To obtain the uncertainty of the gradient measurement, we require the variance of the estimated concentration difference
$\langle[\delta(c_1-c_2)]^2\rangle$, which we calculate using Eq. \ref{eq:dc_n} and substituting 
the previously calculated noise power spectra $S_{n_jn_l}(0)$. %(0)=\langle\delta n_j\delta n_l\rangle$. 
For an averaging time $\tau$,
we obtain for the variance of the inferred concentration difference between two receptors 
separated by a distance $r$:
\begin{eqnarray}
\langle[\delta(c_1-c_2)]_\tau^2\rangle&=&
\frac{(\bar c_1+K_D)^4}{K_D^2\tau}S_{n_1n_1}(0)+\frac{(\bar c_2+K_D)^4}{K_D^2\tau}S_{n_2n_2}(0)\\
&-&\frac{(\bar c_1+K_D)^2(\bar c_2+K_D)^2}{K_D^2\tau}[S_{n_1n_2}(0)+S_{n_2n_1}(0)]\nonumber\\
&=&\frac{2\bar c_1}{k_+(1-n_1)\tau}+\frac{2\bar c_2}{k_+(1-n_2)\tau}+\frac{\bar c_1+\bar c_2}{\pi D\tau}\left(\frac{1}{s}-\frac{1}{2r}\right)\label{eq:var_diff}.
\end{eqnarray}
Assuming that the true gradient is shallow, $\bar c_1\approx \bar c_2$, to estimate the variance in Eq. \ref{eq:var_diff} 
we set $\bar c_1=\bar c_2=c_0$ and $\bar n_1=\bar n_2=n$ for identical receptors in equilibrium, 
and obtain 
\begin{equation}
\langle[\delta(c_1-c_2)]_\tau^2\rangle
=\frac{4c_0}{k_+(1-n)\tau}+\frac{2c_0}{\pi D\tau}\left(\frac{1}{s}-\frac{1}{2r}\right).\label{eq:var_diff2}
\end{equation}
%Notice that this result is lower than the variance of the sum of the concentration fluctuations, {\it i.e.} the quantity
%needed for estimating the concentration, 
%\begin{equation}
%\langle[\delta(c_1+c_2)]_\tau^2\rangle/r^2
%=\frac{4c_0}{k_+(1-\bar n)\tau}+\frac{2c_0}{\pi D\tau}\left(\frac{1}{a}+\frac{1}{2r}\right).
%\end{equation}
%for the normalized uncertainty 
%\begin{equation}
%\frac{\langle[\delta(c_1-c_2)]_\tau^2\rangle/r^2}{(c_0/r)^2}
%=\frac{4}{k_+(1-\bar n)c_0\tau}+\frac{2}{\pi Dc_0\tau}\left(\frac{1}{a}-\frac{1}{2r}\right).
%\end{equation}
%Notice that this result is lower than the variance of the sum of the concentration fluctuations, {\it i.e.} the quantity
%needed for estimating the concentration, 
%\begin{equation}
%\frac{\langle[\delta(c_1+c_2)]_\tau^2\rangle/r^2}{(c_0/r)^2}
%=\frac{4}{k_+(1-\bar n)c_0\tau}+\frac{2}{\pi Dc_0\tau}\left(\frac{1}{a}+\frac{1}{2r}\right).
%\end{equation}
%by an amount $\frac{2}{\pi Dc_0r\tau}$, since correlated fluctuations of nearby receptors
%cancel for gradient measurements.\\

%\begin{equation}
%\left.\frac{\langle[\delta(c_1-c_2)]_T^2\rangle/r^2}{(c/r)^2}\right|_{c_1=c_2=c}=
%\frac{4}{k_+(1-n)cT}+\frac{2}{\pi DcT}\left(\frac{1}{a}-\frac{1}{2r}\right)
%\end{equation}

\noindent{\bf Two rings of receptors}\\ 
Following Bialek and Setayeshgar \cite{bialek05}, we next consider the case where a cell is equipped with 
two rings of $m$ receptors, one at each pole. We further assume
that the cell length is large enough so each receptor in Ring 1 is at nearly the same distance $r$ from each
receptor in Ring 2 (Fig. 2). The equation governing receptor $j$ in Ring 1 is
\begin{eqnarray}
-i\omega\delta\hat n_j^{(1)}(\omega)&=&\left[k_+(1-\bar n_1)\frac{i\omega\Lambda}{2\pi Ds}-(k_+\bar c_1+k_-)\right]\delta\hat n_j^{(1)}(\omega)\\
&&+k_+(1-\bar n_1)\frac{i\omega}{4\pi D}\left[\sum_{l\neq j}^{\text{Ring}\,1}\frac{\delta\hat n_l^{(1)}(\omega)}{|\vec x_j-\vec x_l|}+
\sum_{l=1}^{\text{Ring}\,2}\frac{\delta\hat n_l^{(2)}(\omega)}{r}\right]\\
&&+\frac{k_+(1-\bar n_1)\bar c_1}{k_BT}\delta\hat F_j^{(1)}(\omega).
\end{eqnarray}
Summing over all receptors of Ring 1, and using the notation
$m\delta N^{(1)}=\sum_{j=1}^m\delta n_j^{(1)}(\omega)$ and $\delta F^{(1)}=\sum_{j=1}^m\delta F_j^{(1)}(\omega)$, we obtain for Ring 1
\begin{equation}
\frac{mk_BT}{k_+(1-\bar n_1)\bar c_1}\left\{k_+\bar c_1+k_--i\omega[1+\Sigma_1+\frac{k_+(1-\bar n_1)}{4\pi D}\Phi^{(1)}]\right\}\delta N^{(1)}
-\frac{i\omega mk_BT}{4\pi Dr}\delta N^{(2)}=\delta F^{(1)},
\end{equation}
where $\Phi^{(1)}=\sum_{j\neq 1}^m\frac{1}{|\vec x_1-\vec x_j|}$, with a similar expression 
for Ring 2. These equations can be solved for the uncertainty in measuring the concentration difference between the
rings in a similar manner to the two-receptor case. For an
averaging time $\tau$ and $\Phi=\Phi^{(1)}=\Phi^{(2)}$, one obtains for the uncertainty in measuring a gradient using two rings 
of $m$ receptors each
\begin{equation}
\langle[\delta(c_1-c_2)]_\tau^2\rangle=\frac{2\bar c_1}{mk_+(1-\bar n_1)\tau}+\frac{2\bar c_2}{mk_+(1-\bar n_2)\tau}+
\frac{\bar c_1+\bar c_2}{m\pi D\tau}\left(\frac{1}{s}+\frac{\Phi}{2}-\frac{1}{2r}\right).\label{eq:var_diffring}
\end{equation}
Assuming again that the true gradient is shallow, $\bar c_1\approx \bar c_2$, to estimate the variance in Eq. \ref{eq:var_diffring} 
we set $\bar c_1=\bar c_2=c_0$ and $\bar n_1=\bar n_2=n$ for identical receptors in equilibrium, which results in
\begin{equation}
\langle[\delta(c_1-c_2)]_\tau^2\rangle=\frac{4c_0}{mk_+(1-n)\tau}+
\frac{2c_0}{m\pi D\tau}\left(\frac{1}{s}+\frac{\Phi}{2}-\frac{1}{2r}\right).\label{eq:var_diffring2}
\end{equation}
%or alternatively, for the normalized uncertainty 
%\begin{equation}
%\frac{\langle[\delta(c_1-c_2)]_\tau^2\rangle/r^2}{(c_0/r)^2}
%=\frac{2\bar c_1}{mk_+(1-\bar n_1)c_0^2\tau}+\frac{2\bar c_2}{mk_+(1-\bar n_2)c_0^2\tau}+
%\frac{\bar c_1+\bar c_2}{m\pi Dc_0^2\tau}\left(\frac{1}{a}+\frac{\Phi}{2}-\frac{1}{2r}\right).
%\end{equation}

\section{Results and Discussion}
Many types of cells are known to measure spatial chemical gradients directly with high accuracy. 
In particular, {\it Dictyostelium discoideum} is known to measure extremely 
shallow gradients of cAMP important for fruiting body formation \cite{arkowitz99,mato75,haastert07} 
and {\it Saccharomyces cerevisiae} (budding yeast) detects shallow gradients 
of $\alpha$ mating pheromone \cite{segall93}. 
Direct spatial sensing of gradients is also performed by cells of the immune system 
including neutrophils and lymphocytes \cite{zigmond77}, as well as by the large 
marine bacterium {\it Thiovulum majus} \cite{thar03}.
The question arises what are the limits of the accuracy of gradient sensing 
set by chemical diffusion? Recently, we derived fundamental physical 
limits for gradient sensing \cite{endres_PNAS08} using as model cells a perfectly absorbing sphere and a 
perfectly monitoring sphere \cite{berg77}. We found that a perfectly absorbing sphere is superior to a perfectly monitoring sphere 
for both concentration and gradient sensing since the perfectly absorbing sphere avoids
the noise due to remeasuring previously detected particles (Table I).
Consequently, our results for the perfectly absorbing sphere represent the true
fundamental limits of both concentration and gradient sensing by cells.

Our models of the absorbing and the monitoring spheres 
neglect all biochemical reactions, such as particle-receptor binding and downstream signaling events,
which might further increase measurement uncertainty. To study the effects of particle-receptor binding, we
extended a formalism for the uncertainty of concentration sensing, recently developed by Bialek and Setayeshgar 
\cite{bialek05}, to gradient sensing. This formalism uses the 
fluctuation-dissipation theorem to infer the fluctuations of the receptor occupancy (and
hence the accuracy of concentration sensing) from the linear response of the average
receptor occupancy to changes in receptor binding free energies. The effect of 
particle rebinding is included by coupling particle-receptor binding to the diffusion equation \cite{bialek05}, 
leading to correlations in time among the receptors. \\

\noindent{\bf Single receptor for concentration sensing} \\
It is instructive to first review the result for a single receptor without and with coupling to particle
diffusion (corresponding, respectively, to preventing and allowing rebinding of already measured
 particles) \cite{bialek05}. The uncertainty of sensing concentration $c_0$ without rebinding ({\it i.e.}
assuming that upon unbinding the particle is removed from the system) is given by 
\begin{equation}
\langle(\delta c)_\tau^2\rangle=\frac{2 c_0}{k_+(1-n)\tau}\rightarrow\frac{c_0}{2\pi Ds\tau},\label{eq:no_rebind}
\end{equation}
where $k_+$ is the rate constant for binding, $n$ is the average receptor occupancy, and $\tau$ is the averaging time. 
The right hand side of Eq. \ref{eq:no_rebind} is obtained for diffusion-limited binding, in which case  
$k_+(1-n)\rightarrow 4\pi Ds$ with $D$ the diffusion constant and $s$ the receptor dimension.
This is the well-known fundamental limit derived by Berg and Purcell \cite{berg77}.
In contrast, the uncertainty of concentration sensing including particle diffusion and possible rebinding
is given by Bialek and Setayeshgar as \cite{bialek05} 
\begin{equation}
\langle(\delta c)_\tau^2\rangle=\frac{2(1+\Sigma)c_0}{k_+(1-n)\tau}=\frac{2c_0}{k_+(1-n)\tau}+\frac{c_0}{\pi Ds\tau}
\rightarrow\frac{3c_0}{2\pi Ds\tau},\label{eq:with_rebind}
\end{equation}
where $\Sigma=k_+(1-n)/(2\pi Ds)$. 
Comparison of Eq. \ref{eq:no_rebind} and Eq. \ref{eq:with_rebind} shows that the uncertainty of concentration sensing 
by a single receptor is larger by the term $c_0/(\pi Ds\tau)$ when allowing for rebinding 
of already measured particles. For the minimum uncertainty case set by diffusion-limited binding, and 
given by the right hand side of Eq. \ref{eq:with_rebind},
this additional term simply leads to a factor of 3 increase to the fundamental limit derived by Berg and Purcell (Eq. 
\ref{eq:no_rebind}).\\

\noindent{\bf Two receptors for gradient sensing}\\
As derived in Methods (see  Eq. \ref{eq:var_diff2}), we find that the uncertainty of gradient measurement is given by
\begin{equation}
\langle[\delta(c_1-c_2)]_\tau^2\rangle/r^2
=\frac{4c_0}{k_+(1-n)r^2\tau}+\frac{2c_0}{\pi Dr^2\tau}\left(\frac{1}{s}-\frac{1}{2r}\right)\label{eq:resultdc1}.
\end{equation}
As expected, the larger the receptor-receptor separation $r$,
the smaller the uncertainty in the gradient, because of the larger ``lever arm'' between receptors.
Note that the result in Eq. \ref{eq:resultdc1} for the uncertainty in the gradient is independent
of the magnitude of the actual gradient, including the case when there is no real gradient present.
For comparison, the uncertainty of mean concentration measurement is \cite{bialek05}
\begin{equation}
\langle[\delta(c_1+c_2)/2]_\tau^2\rangle
=\frac{c_0}{k_+(1-n)\tau}+\frac{c_0}{2\pi D\tau}\left(\frac{1}{s}+\frac{1}{2r}\right)\label{eq:resultdc2}.
\end{equation}
Analogous to the single receptor case, Eq. \ref{eq:with_rebind}, 
the first term in Eqs. \ref{eq:resultdc1} and \ref{eq:resultdc2} arises from particle-receptor binding kinetics, 
whereas the second term is due to diffusion and includes the effects of possible rebinding of already measured 
particles. Due to the proximity of the receptors, separated by distance $r$, a particle can
unbind one receptor and subsequently rebind the other receptor because of diffusion (see Fig. 1, trajectory 2).
Hence, there is an additional noise component, which actually improves the accuracy of 
gradient measurement (term $\sim -1/(2r)$) due to cancellation with the noise due to rebinding to the same
receptor, but degrades the accuracy of mean concentration measurement (term $\sim +1/(2r)$) 
since rebinding to the other receptor can only increase noise in the estimate of the mean concentration.
This correlated noise was recently also investigated with Monte Carlo simulations \cite{herbie08}.\\

\noindent{\bf Two rings of receptors for gradient sensing}\\
We next consider two rings of receptors, parallel to one another at opposite cell ends a 
distance $r$ apart (Fig. 2). As derived in Methods (see Eq. \ref{eq:var_diffring2}), 
we find that the uncertainty of gradient measurement is given by
\begin{equation}
\langle[\delta(c_1-c_2)]_\tau^2\rangle/r^2
=\frac{4c_0}{mk_+(1-n)r^2\tau}+\frac{2c_0}{m\pi Dr^2\tau}\left(\frac{1}{s}
+\frac{\Phi}{2}-\frac{1}{2r}\right)\label{eq:resultdc3},
\end{equation}
where $m$ is the number of receptors per ring and $\Phi$ is a geometric factor close to unity.  
For comparison, the uncertainty of mean concentration measurement is \cite{bialek05}
\begin{equation}
\langle[\delta(c_1+c_2)/2]_\tau^2\rangle
=\frac{c_0}{mk_+(1-n)\tau}+\frac{c_0}{2m\pi D\tau}\left(\frac{1}{s}
+\frac{\Phi}{2}+\frac{1}{2r}\right)\label{eq:resultdc4}.
\end{equation}
The factor $1/m$ in Eqs. \ref{eq:resultdc3} and \ref{eq:resultdc4} reflects signal averaging by multiple receptors,
which reduces the measurement uncertainty with respect to the case of two receptors. The possibility of rebinding
to other receptors within the same ring leads to correlations among the signals, which are reflected in the
extra term $\Phi/2$ in the rebinding noise.\\

\noindent{\bf Comparison with the perfect monitor and the perfect absorber models}\\
To make comparison to our results for the perfectly absorbing and monitoring spheres (Table I), which
do not include particle-receptor kinetics, we replace $k_+(1-n) c_0$  
by $4\pi Dsc_0$ for the minimum uncertainty case set by 
diffusion-limited binding. To specifically compare with the perfectly absorbing sphere, 
we neglect the second term in Eqs. \ref{eq:resultdc3} and \ref{eq:resultdc4} (thereby neglecting rebinding of 
particles) and obtain for gradient and concentration sensing
\begin{eqnarray}
\frac{\langle[\delta(c_1-c_2)]_\tau^2\rangle/r^2}{(c_0/r)^2}&=&\frac{1}{\pi Da'c_0\tau}\label{eq:diff_limited1}\\
\frac{\langle[\delta(c_1+ c_2)/2]_\tau^2\rangle}{c_0^2}&=&\frac{1}{4\pi Da'c_0\tau}\label{eq:diff_limited2}
\end{eqnarray}
respectively. To specifically compare with the perfectly monitoring sphere we keep both terms in 
Eqs. \ref{eq:resultdc3} and \ref{eq:resultdc4} and obtain for gradient and concentration sensing
\begin{eqnarray}
\frac{\langle[\delta(c_1-c_2)]_\tau^2\rangle/r^2}{(c_0/r)^2}
&=&\frac{1}{\pi Da'c_0\tau}\left[3+s\left(\Phi-\frac{1}{r}\right)\right]\label{eq:diff_limited3}\\
\frac{\langle[\delta(c_1+c_2)/2]_\tau^2\rangle}{c_0^2}
&=&\frac{1}{4\pi Da'c_0\tau}\left[3+s\left(\Phi+\frac{1}{r}\right)\right]\label{eq:diff_limited4}.
\end{eqnarray}
The parameter $a'=ms$ is the combined receptor dimension, ultimately limited by the cell dimension. 
Note that in Eqs. \ref{eq:diff_limited1}  and \ref{eq:diff_limited3} for gradient sensing we
normalized by $(c_0/r)^2$, and in Eqs. \ref{eq:diff_limited2}  and \ref{eq:diff_limited4} for 
concentration sensing we normalized by $c_0^2$ in order to use the same notation 
as Table I and Ref. \cite{endres_PNAS08}.

As a result, for $r\geq s$, {\it i.e.} receptor separation larger than receptor size,
the measurement uncertainty with rebinding (Eqs. \ref{eq:diff_limited3} and \ref{eq:diff_limited4}) 
is always larger than the measurement uncertainty without rebinding
(Eqs. \ref{eq:diff_limited1} and \ref{eq:diff_limited2}) for both gradient and concentration sensing. 
Hence, the absorber is superior to the monitor even when 
receptor binding kinetics are explicitly included in line with our previous finding (Table I). Specifically,
for diffusion-limited binding, the dominant effect of particle rebinding (Eqs. \ref{eq:diff_limited3} 
and \ref{eq:diff_limited4}) is simply an increased numerical prefactor, also in line with our results
for the perfect absorber and perfect monitor models.\\

In conclusion, we found that the accuracy of concentration and gradient measurement without
ligand rebinding is higher than the accuracy with rebinding, confirming
the superiority of the absorber over the monitor \cite{endres_PNAS08}.
Our model of two coaxial rings qualitatively resembles the polar clusters found abundantly in
bacteria and archaea \cite{gestwicki00}. Hence, our model may be directly suitable 
for describing the concentration sensing by these organisms and possibly also for
oxygen-gradient sensing by the bacterium {\it Thiovulum majus} \cite{thar03}. 
Furthermore, a number of mechanistic models for gradient sensing and chemotaxis by eukaryotic cells
have addressed the important questions of cell polarization, signal amplification, and 
adaptation \cite{meinhardt99,skupsky05,narang05,levine06,krishnan07,onsum07,otsuji07}, 
cell movement of individual cells \cite{dawes06,dawes07}, cell aggregation \cite{palsson97},
as well as sensing of fluctuating concentrations \cite{bialek05,goodhill99,wylie06,herbie08}.
Our results on the accuracy of gradient sensing complement these models, and
may ultimately help lead to a comprehensive description of eukaryotic chemotaxis \cite{iglesias08}.

\begin{acknowledgments}
RGE acknowledges funding from the
Biotechnology and Biological Sciences Research Council grant BB/G000131/1 
and the Centre for Integrated Systems Biology at Imperial College (CISBIC). 
NSW acknowledges funding from the Human Frontier Science Program (HFSP) 
and the National Science Foundation grant PHY-0650617.
\end{acknowledgments}

\end{singlespace}
\end{document}